# THE TRANSACTION AS A QUANTUM CONCEPT

**Leonardo Chiatti**

AUSL VT Medical Physics Laboratory
Via Enrico Fermi 15 – 01100 Viterbo (Italy)
fisica1.san@asl.vt.it

**ABSTRACT**

This essay intends to present a novel approach to the concept of "transaction" in quantum physics. The central ideas of this approach were outlined by this author in two essays in the 1990s (1, 2), while a more detailed treatment was published in volume in 2005 (3). Breaking with Cramer's original theory, the transaction is not connected to the simultaneously retarded and advanced spacetime propagation of classical fields, as in the spirit of Wheeler-Feynman electrodynamics. Instead, the transaction is seen as an archetypal structure intrinsic within the quantum formalism. The present approach is advantageous in that, while preserving the essential point of Cramer's theory, is also fully consistent with the standard quantum formalism. In particular, it has the advantage of avoiding the introduction of elements which are completely extraneous to quantum formalism (such as the propagation of real physical waves in four-dimensional spacetime, the phase difference between offer and confirmation waves which is necessary for the elimination of "tails", the echoing mechanism, etc.) and which have led to misunderstandings, such as Maudlin's objection. Furthermore, this approach elucidates the relationship between transactional mechanism and block universe, implicate order, and acausality.

## 1. INTRODUCTION

Transaction is normally associated with a particular interpretation of quantum formalism, the so called Transactional Interpretation (TI) introduced by Professor J.G. Cramer in the 1980s (4,5,6). All interpretations of quantum formalism introduce elements extraneous to the formalism itself. These elements (which can be broadly associated with Bell's "beables") are introduced with the aim of making quantum formalism understandable by placing it in the context of a given ontology.

In TI the basic quantum process - which includes the preparation of the initial quantum state, the detection of the final quantum state and the connection between these two events - is described through the introduction of the following beables:

- A retarded offer wave propagating from the preparation event towards the possible detection events;

- A source of the aforementioned wave, essentially consisting of the preparation event;

- An advanced confirmation wave propagating from the detection event to the various possible preparation events;

- A source of the aforementioned wave, consisting of the actualized detection event.

Transaction is the feedback (or handshake) procedure that closes the loop between the effectively realized preparation and detection events. Such a procedure is clearly non-local, albeit in principle relativistically invariant. It can take place in one step or become completed in a finite number of echoes in a sort of pseudotime. Apart from this latter detail, this scheme resembles that of Wheeler-Feynman's electrodynamics (7,8), which is a pre-quantum or classical theory. Cramer's aim therefore is to derive the basic quantum process by using classical pre-quantum concepts, especially in those aspects that deviate from classical theory the most: the quantum localization and the randomness of the final (initial) state with an assigned initial (final) state.

It must be noted that the basic quantum process is, according to TI, a dynamic process taking place in spacetime. The offer and confirmation waves propagate in spacetime and the preparation and detection events are localized in spacetime. In the prevalent terminology during the 1980's, the TI can be defined as a non-local realist interpretation.

In order to achieve his aim, Cramer attributes to his beables certain patterns of behavior. Beside the above mentioned echoing mechanism, Cramer assumes that the confirmation wave is out of phase with respect to the offer wave, so that the tails (those regions of the offer wave that follow the detection, and those regions of the confirmation wave that precede the preparation) cancel each other via destructive interference. Cramer also assumes that the two waves can be represented as a superposition of the eigenfunctions of the pertinent observables, with coefficients that are exactly those same necessary to ensure compliance with Born's rule. This latter requirement is not altogether obvious and leads to difficulties that have been clearly evidentiated by Maudlin (9) and discussed by Berkovitz (10) and Kastner (11).

This essay intends to show that once Cramer's time-symmetrical point of view is adopted, the formal structure of the transactional loop is already expressed by the current quantum formalism, making the introduction of the beables unnecessary. As a matter of fact, transaction is already present in formalism and is fully consistent with it, so that special interpretations such as TI are unnecessary. Given this, Cramer's solution should be regarded not so much as the proposal of a new interpretation, but rather as the discovery of a formal structure underlying the quantum theory that has previously gone unrecognized or at least underestimated.

The necessary step seems to be rather the adoption of an ontology of the basic quantum process strictly consistent with formalism and its current metatheoretical interpretation. Therefore, only the events of preparation of the initial state (creation) and of detection of the final state (annihilation or destruction) are assumed to be real physical events. The connection between these two events is ensured by their common origin in an atemporal and aspatial background outside spacetime. The process amplitudes are a mathematical fiction suited to calculate the statistics of the connection itself. Accordingly, the virtual processes obtained by expanding these amplitudes into partial amplitudes are also mathematical fiction. The acausality of the basic quantum porcess is intrinsic to its emergence from a background which is outside the spacetime. The dynamics of the basic quantum process is in reality *a statics*: the forward and backward transition amplitudes (which substitute here the Cramer's offer and confirmation waves respectively) are not associated with any propagation into spacetime.

Strictly speaking, only the events of creation/annihilation associated with an actual localization in the spacetime can be connected to the pointevents of the relativistic chronotope; generally, however, such events cannot be represented in spacetime. Transaction is therefore an archetypal structure, functioning as a bridge between two levels of physical reality: one is synchronic (a sort of *arché*[1]) and the other diachronic. This bridge, of course, cannot be exclusively represented on the diachronic level by means of purely diachronic mechanisms, albeit time-symmetrical, such as those envisioned in traditional TI.

Research along these lines should focus on the definition of a theory postulating the emergence of physical reality from the *arché*. A complete derivation of quantum formalism, of quantum dynamics and of transactional mechanism should be a particular aspect of such general theory. This is actually the program that Bohm had already sketched out when dealing with clarifying the relationship between *implicate* and *explicate order* (12). In what follows, I will confine myself to illustrating some aspects of the relationship between the transactional mechanism and quantum statistics while I remind to ref. (3) for an extensive presentation of a possible example of such a derivation.

---

[1] Greek for "origin" or "first cause".

The outline of the present essays is as follows: Section 2 contains an overview of transaction as a structure built in the quantum formalism. Section 3 deals in details with some logical elements of the problem. Section 4 shows the derivation of quantum statistics. Owing to the limits of this essay dynamics laws are not worked out; the reader is referred to another work (3) for this subject. Section 5 contains a simple justification of the well-known correspondence between thermodynamics and quantum mechanics, which is here taken to be a rule of the *unfolding* of the transactional loop.

## 2. TRANSACTIONS: AN OVERVIEW

We assume that the only truly existent "thing" in the physical world are the events of creation and destruction (or, if one prefers, physical manifestation and demanifestation) of certain qualities. In the language of QFT these events are the "interaction vertices", while the different sets of manifested/demanifested qualities in the same vertex are the "quanta".

As an example, in a certain vertex a photon (E, **p**, **s**) can be created, where E is the (created) energy of the photon, **p** is the (created) impulse of the photon and **s** is the (created) spin of the photon. In a subsequent event this photon can be absorbed and this corresponds to a packet of properties (-E, -**p**, -**s**) where –E is the (absorbed) energy of the photon, - **p** is the (absorbed) impulse of the photon and – **s** the (absorbed) spin of the photon; (-E, -**p**, -**s**) is the absorbed, i.e. destroyed, photon. It is assumed that the first event chronologically precedes the second, and that E is positive. A process is therefore being described in which a positive quantity of energy is created and then destroyed. In an absolutely symmetrical manner, it can be said that we are describing a process which proceeds backwards in time and during which a share of negative energy –E is firstly created (in the second event) and then destroyed (in the first event). The two descriptions are absolutely equivalent.

There are two reasons for which, in the process described, one cannot have E < 0, i.e. the propagation of a positive energy photon towards the past. A photon with E > 0 which retropropagates towards the past, yielding energy to the atoms of the medium through which it travels (such as, say, a photon X which ionizes the matter through which it travels), is seen by an observer proceeding forward in time as a photon with E < 0 which *absorbs* energy from the medium (13). This photon would be spontaneously "created" by subtracting energy from the medium, through a spontaneous coordination of uncorrelated atomic movements which is statistically implausible, and has never been observed experimentally.

From a theoretical point of view, the probability of the occurrence of a creation/destruction event for a quantum $Q$ in a point event $x$ is linked to the probability amplitude $\Psi_Q(x)$, which can be a spinor of any degree. Each component $\Psi_{Q,i}(x)$ of this spinor satisfies the Klein-Gordon quantum relativistic equation $(-\hbar^2 \partial^\mu \partial_\mu + m^2 c^2) \Psi_{Q,i}(x) = 0$, where $m$ is the mass of the quantum[2]. At the non-relativistic limit, this equation becomes a pair of Schrödinger equations (14):

$$-\frac{\hbar^2}{2m} \Delta\Psi_{Q,i}(x) = i\hbar\, \partial_t\, \Psi_{Q,i}(x) \qquad (1)$$

---

[2] It is necessarily positive, because it is a measure of the energy which must be released in order to create the quantum; a physical entity can certainly not be created in the vacuum by subtracting energy from it.

$$-\frac{\hbar^2}{2m} \Delta \Psi^*_{Q,i}(x) = -i\hbar\, \partial_t \Psi^*_{Q,i}(x) \qquad (2)$$

The first equation has only retarded solutions, which classically correspond to a material point with impulse **p** and kinetic energy $E = \mathbf{p}\cdot\mathbf{p}/2m > 0$. The second equation has only advanced solutions, which correspond to a material point with kinetic energy $E = -\mathbf{p}\cdot\mathbf{p}/2m < 0$. Thus there are no true causal propagations from the future.

One may wonder whether eq. (2) can lead to hidden advanced effects. The answer to this question is affirmative. To understand this topic, one ought to reconsider the photon example seen above. The creation of the $E > 0$ energy followed by its subsequent absorption and, conversely, the creation of a $-E < 0$ energy preceded by its destruction are clearly two different descriptions of the same process. This however is true so long as the interaction vertices are considered i.e., according to the here adopted ontology, the true substance of the physical world[3].

From the point of view of the dynamical laws for the probability amplitudes of these events, matters are quite different, however. The creation of quality $Q$ is associated with the initial condition for $\Psi_{Q,i}(x)$ in eq. (1); the destruction of quality $Q$ is associated with the "initial", actually the final, condition for $\Psi^*_{Q,i}(x)$ in eq. (2). In general, however, the two conditions are different and therefore generate different solutions for the two equations, which are not necessarily mutual complex conjugates.

It is a fundamental fact that the destruction event is not described by eq. (1) and that the creation event is not described by eq. (2); this remains true even if the Hamiltonians of interaction with the remaining matter are introduced into the two equations. Thus, at the dynamic laws level, the process of the creation and destruction of $Q$ is completely described solely by the loop:

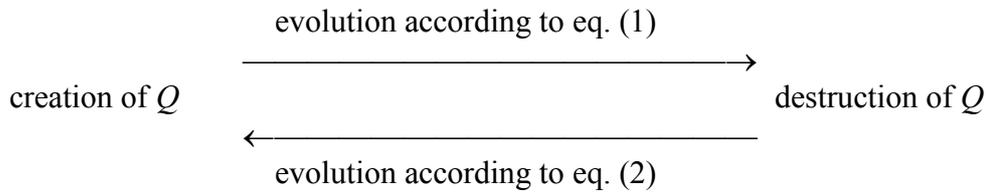

and not only by the upper or lower half-loop.

More generally speaking, we shall have at $t = t_1$ the event of the creation-destruction of a quality $Q$ ( $|Q><Q|$ ) and at $t = t_2$ the event of the creation-destruction of a quality $R$ ( $|R><R|$ ). These two processes will be linked by a time evolution operator $S$ according to the ring:

---

[3] As we will see in the following, the term "interaction vertex" has to be intended as referred to an elemental event of interaction between real particles, or more generally to an elemental event involving an objective "reduction" of the quantum state.

$$\begin{array}{ccc} |Q\rangle & \langle Q| & t=t_1 \\ S & |R'\rangle & \\ & \diagdown\diagup & \\ & \diagup\diagdown & \\ S^+ & |Q'\rangle & \\ |R\rangle & \langle R| & t=t_2 \end{array} \qquad (3)$$

In other words, $|Q\rangle$ is transported from $S$ into $|Q'\rangle$ and projected onto $\langle R|$, $|R\rangle$ is transported by $S^+$ into $|R'\rangle$ and projected onto $\langle Q|$. The amplitudes product:

$$\langle R|S|Q\rangle \langle Q|S^+|R\rangle = |\langle R|S|Q\rangle|^2$$

is immediately obtained, which is the probability of the entire process. If quality $Q$ is constituted by a complete set of constants of motion then $R = Q$ and this is the type of process which can describe the propagation of a photon-type quantum[4], otherwise it is the generic process of the creation of a quality $Q$ causally linked (by means of $S$) to the destruction of a quality $R$. Moving to the representation of the coordinates, by substituting bras and kets with wavefunctions, we once again obtain as a particular case the result already seen with the non-relativistic expressions (1), (2). The ring process described above will be called *transaction*; within quantum formalism, it plays the role that in Cramer's TI is reserved to the response of confirmation waves to an offer wave. The transaction exists if the propagation $S^+$ is as "real" as the $S$; to ascertain this, one must see whether experimental situations exist in which the initial condition $|Q\rangle$ and the final one $\langle R|$ are connected in a nonlocal way. As is well known, the answer is affirmative: the phenomena EPR and GHZ (15) are highly valid examples of quantum-mechanical predictions which violate locality. In the case of EPR phenomena there is now confirmed solid experimental evidence (16-19).

Some remarks.

1) Quantum nonlocal effects do not follow from the forward propagation $S$ only. However, the existence of such effects is well established at present, which implies that another factor is involved. The existence of advanced propagation $S^+$ is a logical inference since it is reflected in the Born probability and in the probability amplitudes.

2) In the same hypothesis, i.e., considering only the forward propagation, one cannot explain the destruction of the "quantum state" as a phenomenon that takes place at a defined instant $t_2$. Well known paradoxes, such as that of Schrödinger's cat (20), derive from this.

---

[4] We can also describe this process as the propagation of a single quality. In conventional TI this would be a singletransaction with probability of unity

3) From an algebraic point of view, the transactional ring is a sort of identity operator, because $SS^+ = S^+S = 1$ and the qualities $Q$, $R$ are simultaneously created and destroyed. One has the impression that every quantum process (therefore all matter) and time itself are emitted from an invariant substratum and re-absorbed within it. This substratum cannot be observed because it is invariant and outside of space and time, which originate from it: a sort of "motionless motor".

    The transactional ring may be described by circular inference rules which establish a self-consistency rather than, as in traditional logic, by linear inference rules which establish a deduction. A circular logic, not a linear one, is applied to a ring; it appears therefore that individual quantum processes are self-generated in a non-causal manner through an extra-spacetime mechanism.

    Self-generation implies acausality. At the same time, however, the creation/destruction events that occur at $t = t_1$, $t = t_2$ take place because of interactions with other rings. This implies the existence of rules on how the rings connect and therefore the acausality does not turn into complete arbitrariness. This appears to be a natural and entirely convincing explanation of the simultaneous presence of causality and acausality in quantum processes.

4) The energy is propagated only in one time direction, and the causal effects thus proceed from the past towards the future.

To sum up, the two extreme events of a transaction correspond to two reductions of the two state vectors which describe the evolution of the quantum process in the two directions of time. They constitute the "**R** processes" (**R** stands for *reduction*) of the Penrose terminology (21-23) and, from our perspective, are the only real physical processes. They are constituted of interaction vertices in which real elementary particles are created or destroyed; these interactions are not necessarily acts of preparation or detection of a quantum state in a measurement process.

The evolution of probability amplitudes in the two directions of time constitutes, in Penrose terminology, a **U** process (where **U** stands for *unitary*). From our ontological viewpoint, **U** processes are not real processes: both the amplitudes and the time evolution operators which act on them are mathematical inventions whose sole purpose is to describe the causal connection between the extreme events of the transaction, i.e. between **R** processes. This connection is possible because the two events derive from the transformation of the same aspatial and atemporal "substratum". As a specific consequence of this assumption, all virtual processes contained in the expansion of the time evolution operator are deprived of physical reality.

According to this approach, therefore, the history of the Universe, considered at the basic level, is given neither by the application of forward causal laws at initial conditions nor by the application of backward causal laws at final conditions. Instead, it is assigned as a whole as a complete network of past, present and future **R** processes. Causal laws are only rules of coherence which must be verified by the network and are *per se* indifferent to the direction of time; this is the so called *block universe* in Putnam and Rietdijk's conception (24, 25).

A transaction that begins with the creation of quality $q$ and ends with the destruction of quality $r$ can be represented simply by the form $q^+ r^-$. For clarity's sake, however, we will use here also another symbolism, the one introduced by Costa de Beauregard (26-29). According to it, the creation $q^+$ of the property $q$ is indicated with $| q )$, while the destruction $r^-$ of property $r$ is indicated with $( r |$. The transaction $q^+ r^-$ is then denoted as $( r | q )$. We remark that a symbol as, e.g., $| q )$ represents an operator belonging to a suitable algebraic structure and should be not confused with the associated ket $| q >$ which is instead an amplitude.

As an example of a transactional network, let us consider a process well known in QFT, constituted by the decay of a microsystem, prepared in the initial state 1, into two microsystems 2, 3 which are subsequently detected. The preparation consists of the destruction of quality 1 [which we shall indicate with ( 1 |] which closes the transaction which precedes it, and of the creation of quality 1 [which we shall indicate with | 1 )] which opens a new transaction. It will be represented by the form | 1 )( 1 |. The decay consists of the destructions of qualities 2, 3 which close the transaction that began with the preparation, and of the creations of qualities 2, 3 which open a new transaction which will be closed with the detection of microsystems 2, 3. It will be represented by the form [| 2 )| 3 )] [( 2 |( 3 |].

The detection of microsystems 2, 3 will be constituted by the destructions of qualities 2, 3 which close the transaction that began with the decay, and by the creations of qualities 2, 3 which open subsequent transactions. It will be represented by the interaction events | 2 )( 2 |, | 3 )( 3 |.

The double transaction described here corresponds to the process usually associated with the probability amplitude < 2, 3 | S | 1 >.

Another example is Young's classical double slit experiment. The preparation of the initial state of the particle can be represented by the form | 1 )( 1 |, following the same reasoning as in the previous case. Instead of the decay, here we have the encounter with slits 2 and 3, i.e. the interaction between a particle and a double slit screen represented by [| 2 )| 3 )] [( 2 |( 3 |]. Instead of the detection of the two particles created in the decay, here we have the sole event of the detection of the particle on the second screen at a certain position 4, i.e. : | 4 )( 4 |. Two transactions are involved: the first starts with the preparation of the particle and ends with its interaction with the first screen; the second begins with this second interaction and ends with the interaction of the particle with the second screen. The latter interaction then constitutes the beginning of the following transaction. The process is that which corresponds to the probability amplitude < 4 | S | 1 >.

We note that the forward time evolution of the amplitudes, represented with $S| 1 >$, contains both the kets | 2 >, | 3 >; nevertheless, processes relating to the passage through the individual slit $a$ (where $a$ = 2, 3) do not exist. Such processes would require an intermediate event represented by| $a$ )( $a$ |, which effectively does not take place. It is in this sense that processes that can be associated with compound probability amplitudes < 4 | S | $a$ > < $a$ | S | 1 > are "virtual" and not real. The process of the crossing of one of the two slits becomes real when the other slit is closed.

## 3. THE PHYSICAL UNIVERSE AS NETWORK OF R MICROEVENTS

Let us focus on some of the concepts summarily reviewed in the previous Section. A single event separating two transactions is denoted as a product of destructions (which close the preceding transaction) and creations (which open the subsequent transaction). For example, the event $A$ = | $s$ ) ( $u$ | consists of the destruction of $u$ taking place in time $t_A$ - $\varepsilon$, and the creation of $s$ taking place in time $t_A$ + $\varepsilon$, where $t_A$ is the instant in which the event $A$ occurs and $\varepsilon \to 0$.

Also, the symbol $A$ = | $s$ ) | $u$ ) denotes the event $A$ consisting of the creation of $u$ simultaneous with the creation of $s$, and so forth.

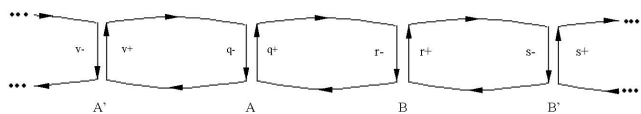

Fig. 1

Two or more transactional rings can be hooked to each other in a succession or sequence as shown in Fig. 1. As one can see, the generic event $A = | q ) ( q |$ closes, with its component $q^-$, the preceding transaction (whose initial event can possibly be placed to the past infinity $t = -\infty$); with its component $q^+$ it opens up the subsequent transaction (whose final event can possibly be placed to the future infinity $t = +\infty$). Therefore, the event $A$ joins the transactions $v^+ q^-$ and $q^+ r^-$. In a similar way, the event $B = | r ) ( r |$ joins the transactions $q^+ r^-$ and $r^+ s^-$, and so on.

Two or more such sequences can cross each other, intersecting in a certain event. In Fig. 2 the event $O = (w^+, w^-, z^+, z^-)$ belongs to the sequence C1 through operations $w^-, w^+$ which respectively close a transaction in C1 and open the following one in C1; the same event also belongs to the sequence C2 through operations $z^-, z^+$ which respectively close a transaction in C2 and open the following one in C2. Several sequences, by intersecting each other, will create a net provided with edges and vertices: a net of interconnected physical events. The physical world consists of the set of such nets. The sequences are real (not virtual) physical processes. Each event (that is, each vertex of the net) represents an *interaction*. If it is placed at the intersection of two or more sequences then it represents the real (not virtual) interaction between the real physical processes embodied by these sequences, as the decay $1 \to 2 + 3$ considered above; otherwise the interaction is really a null interaction, as in quantum measurement with a negative result. For example, let us consider a Young's interferometer; when the particle's passage through the first slit is not detected, this means that the particle went through the second slit. With $q$ = "position of the second slit", the event $q^+ q^-$ is then generated; this event is a null interaction.

We remark that, if an algebra exists which reflects the structure of the interaction in $A$ and $| q ) = | r ) | s )$ is a well formed formula of this algebra, then the event of interaction $A = | q ) ( q |$ can actually take the form $A = | r ) | s ) ( q |$, as in the examples discussed in the previous Section.

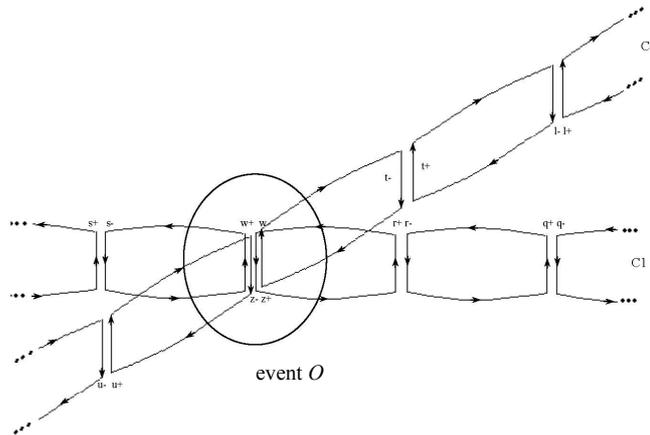

Fig. 2

## 4. THE STATISTICS OF TRANSACTIONS

Each transaction is a self-connection of the aspatial and timeless physical vacuum. This process can be described as the concurrence of two distinct transformations. In the first transformation, the vacuum divides into two pairs of opposites: $q^+$, $q^-$ and $r^+$, $r^-$ ; the first pair forms "event $A$", the second pair forms "event $B$" :

$$A = |\,q\,)\,(\,q\,|\,, \qquad B = |\,r\,)\,(\,r\,| \qquad .$$

The expression $|\,q\,)\,(\,q\,|$ can be considered as representative of an infinite set of loops as that depicted in Fig. 3a (oriented self-connections of the event $A$). Analogously, the expression $|\,r\,)\,(\,r\,|$ can be considered as representative of an infinite set of loops as that depicted in Fig. 3b (oriented self-connections of the event $B$)[5].

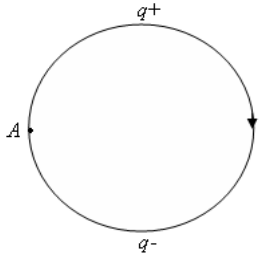 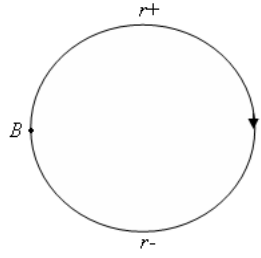

Fig. 3a  Fig. 3b

The second transformation is the real generation of the transaction having the events $A$ and $B$ as its extremities. It consists in the breaking of $N$ loops of the first group and $M$ loops of the second group. For what concerns the first group, the free ends of the broken lines are connected to the event $B$ (Fig. 3c); analogously, the free ends of the broken lines of the second group are connected to the event $A$ (Fig. 3d). Finally, $A$ and $B$ events are connected by means of $M+N$ lines oriented from $A$ towards $B$, and $M+N$ lines having opposite verse.

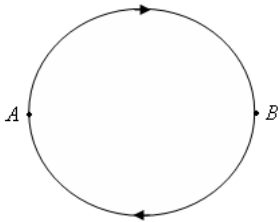 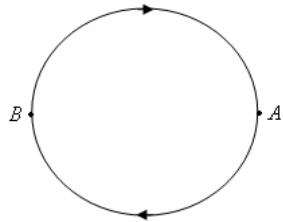

Fig. 3c  Fig. 3d

We postulate that each line of the first (second) group entering into the points $A$, $B$ joints with every line of the same group exiting from the same point, so generating a closed loop. In this way $N^2$ closed loops are formed with lines belonging to the first group, and $M^2$ closed loops are formed with lines belonging to the second group. If one assumes the *a priori* equiprobability of all these self-connections of vacuum, the global statistical weight of the process is expressed by $N^2 + M^2$.

---

[5] Graphs 3a, 3b resemble Feynman graphs of vacuum-vacuum coupling; however, these last present at least two vertices. What we are really describing here is the emergence of the quality $q$ from an aspatial and timeless vacuum. Let us remark that $q$ can be (or can include) also a spacetime position. Therefore, the vacuum we refer to is actually the source of spacetime, rather than the state of minimum energy of a field defined on the spacetime. It should not be confused with the vacuum as intended in Quantum Field Theory (which underlies to Feynman graphs calculation). For an elucidation of the relation between these two notions of physical "vacuum" we remind to ref. (3).

These self-connections are acausal. Let us now consider a set of transactions having the same event at one extremity (we say, $A$) and the other extremity (we say, $B$) freely varying. Therefore the created quality $q$ is exactly determined, while the destroyed quality $r$ can assume a value selected in the set $\{r_1, r_2, ..., r_L\}$.

In other terms:

$$A = |q)(q|, \qquad B = B_i = |r_i)(r_i| \qquad i = 1, 2, ... L .$$

We thus have $L$ possible distinct $q^+ r_i^-$ transactions ($i = 1, 2, ... L$), each of these involving $N(i)$ loops of the first group and $M(i)$ loops of the second group. For the reasons detailed above, the statistical weight of the $i$-th transaction is $[N(i)]^2 + [M(i)]^2$. Thus, the probability that the effectively bootstrapped transaction coincides with the $k$-th process amounts to:

$$P(q^+ r_k^-) = \{[N(k)]^2 + [M(k)]^2\} / \sum_i \{[N(i)]^2 + [M(i)]^2\}$$

$i, k = 1, 2, ..., L$.

Let us introduce, therefore, the "probability amplitude" of the $k$-th transaction:

$$\langle r_k | q \rangle = Z^{-1/2} \{[N(k) + M(k)] + i [N(k) - M(k)]\} \quad , \tag{4}$$

where $\qquad Z = 2 \sum_j \{[N(j)]^2 + [M(j)]^2\} \quad .$

One immediately obtains that $P(q^+ r_k^-) = |\langle r_k | q \rangle|^2$.

The transaction in which the quality $r_k$ is created rather than destroyed, and the quality $q$ is destroyed rather than created, differs from the transaction in question solely by the exchange of events $A$ and $B$. This is equivalent to exchanging $N(i)$ and $M(i)$. Thus, the amplitude of this second transaction is:

$$\langle q | r_k \rangle = Z^{-1/2} \{[M(k) + N(k)] + i [M(k) - N(k)]\} \quad , \tag{5}$$

where $Z$ maintains the same value. Comparing the expressions of the two amplitudes one thus has that:

$$\langle q | r_k \rangle = \langle r_k | q \rangle^* \quad .$$

The probability of this inverse transaction is therefore $P(r_k^+ q^-) = |\langle q | r_k \rangle|^2$. It is equal to that of the direct transaction: $P(r_k^+ q^-) = P(q^+ r_k^-)$.

We can observe that absorbing the factor $Z^{-1/2}$ in $M(k)$, $N(k)$, relations (4), (5) can be inverted so obtaining:

$$M(k) = \text{Re}[\langle q \mid r_k \rangle (1 - i)]/2 \qquad (4')$$

$$N(k) = \text{Re}[\langle r_k \mid q \rangle (1 - i)]/2 \qquad (5')$$

Eqs. (4'), (5') are extremely important, because they connect some transempirical properties of the vacuum (as $M(i)$ and $N(i)$ are) with the usual quantum formalism.

Let us return to the direct transaction ( $r \mid q$ ) we have considered at the beginning of this section, associated with a probability amplitude $\langle r \mid q \rangle$. Let us suppose that, in the graph in Fig. 3a, the set of self-connections (loops) $A \to A$ is subdivided into a certain number of subsets, say, two for the sake of example. Therefore, the transaction involves $N_1$ loops belonging to the first subset, and $N_2$ loops belonging to the second subset. Obviously $N = N_1 + N_2$.
We assume the same partition holds for the loops $B \to B$ of the second group (Fig. 3b). Therefore, we have $M_1$ loops belonging to the first subset and $M_2$ loops belonging to the second subset. Again, it is $M = M_1 + M_2$.

One can therefore write:

$$\langle r \mid q \rangle =$$

$$Z^{-1/2} \{ [N+M] + i [N-M] \} =$$

$$Z^{-1/2} \{ [N_1 + N_2 + M_1 + M_2] +$$

$$\qquad i [N_1 + N_2 - M_1 - M_2] \} =$$

$$Z^{-1/2} \{ [N_1 + M_1] + i [N_1 - M_1] \} +$$

$$Z^{-1/2} \{ [N_2 + M_2] + i [N_2 - M_2] \} =$$

$$= \langle r \mid q \rangle_1 + \langle r \mid q \rangle_2 \ .$$

Consequently:

$$P(q^+ r_k^-) = |\langle r \mid q \rangle|^2 =$$

$$|\langle r \mid q \rangle_1|^2 + |\langle r \mid q \rangle_2|^2 +$$

$$\langle r \mid q \rangle_1 \langle r \mid q \rangle_2^* + \langle r \mid q \rangle_2 \langle r \mid q \rangle_1^* =$$

$$Z^{-1} \{ [N_1^2 + M_1^2] + [N_2^2 + M_2^2] + 2[N_1 N_2 + M_1 M_2] \} \ . \qquad (6)$$

The appearance of the last term constitutes the well-known phenomenon of self-interference. With reference to the popular example of the double slit, the index 1 (2) represents "particle paths" crossing the slit 1 (2), while $q$ and $r$ represent respectively the initial and the final state of the particle.
The origin of this phenomenon is far from mysterious. We remember that $A$ and $B$ events are connected by means of $M+N$ lines oriented from $A$ towards $B$, and $M+N$ lines having opposite verse.

Clearly, is well possible that a line of the first group belonging to the subset 1 joints with a line of the first group belonging to the subset 2, etc.
There are four possibilities:

> 1) Through *A* and *B* a half-loop of the subset 1 is connected to an inverse half-loop of the subset 1. First term of Eq. (6).
>
> 2) Through *A* and *B* a half-loop of the subset 2 is connected to an inverse half-loop of the subset 2. Second term of Eq. (6).
>
> 3) Through *A* and *B* a half-loop of the subset 1 is connected to an inverse half-loop of the subset 2. Last term of Eq. (6).
>
> 4) Through *A* and *B* a half-loop of the subset 2 is connected to an inverse half-loop of the subset 1. Last term of Eq. (6).

Naturally, if only one subset of connections is realized in the actual transaction, self-interference vanishes because the sum of the amplitudes is then confined to the sole term actually present (1 or 2). This situation occurs in the case, for example, of the double slit experiment when only one slit is left open and the other is closed.

An important case is that in which the subdivision into subclasses takes place on the basis of the results which one would have had by subdividing the transaction into two distinct <u>successive</u> transactions. To understand this topic, one must firstly observe that the probability of the process consisting of the two successive transactions ( $r_k \mid q$ ) and ( $s \mid r_k$ ) is obviously given by the product of the probabilities of the two transactions, i.e.:

$$P(q^+ r_k^-) \times P(r_k^+ s) =$$

$$|\langle r_k \mid q \rangle|^2 \times |\langle s \mid r_k \rangle|^2 =$$

$$|\langle r_k \mid q \rangle \langle s \mid r_k \rangle|^2.$$

The rule whereby the probability of the process is given by the modulus squared of the amplitude (Born rule) therefore remains valid in this case, too, provided that the amplitude of the process ( $s \mid r_k$ ) ( $r_k \mid q$ ) is defined in the following manner:

$$\langle s \mid r_k \rangle \langle r_k \mid q \rangle.$$

Let us now consider a transaction ( $s \mid q$ ) during which the quality $q$ is created and the quality $s$ is destroyed. It can be imagined as the simultaneous actuation of all the processes:

$$(s \mid r_k)(r_k \mid q) \; ; \qquad k = 1, 2, ..., L \; .$$

In other words, each transactional half-loop connecting the events $\mid q \rangle \langle q \mid$ and $\mid s \rangle \langle s \mid$ can be seen as the succession of two edges: one connecting $\mid q \rangle \langle q \mid$ to $\mid r_k \rangle \langle r_k \mid$ , the other connecting $\mid r_k \rangle \langle r_k \mid$ to $\mid s \rangle \langle s \mid$. The transactional half-loops can therefore be subdivided into subsets corresponding to the various $r_k$ ($k = 1, 2, ..., L$) values of $r$.

As we have seen, the total transaction amplitude can thus be expressed as the sum of the partial amplitudes relating to the various subsets:

$$\langle s | q \rangle = \sum_k \langle s | r_k \rangle \langle r_k | q \rangle .$$

This result allows the *ket* to be defined:

$$| q \rangle = \sum_k | r_k \rangle \langle r_k | q \rangle .$$

The transaction probability amplitude is thus obtained by "left-multiplying" this expression by $\langle s |$. Alternatively, one can define the *bra*:

$$\langle s | = \sum_k \langle s | r_k \rangle \langle r_k | .$$

The transaction probability amplitude is thus obtained by "right-multiplying" this expression by $| q \rangle$. It obviously follows from the relations:

$$\langle r_k | q \rangle = \sum_i \langle r_k | r_i \rangle \langle r_i | q \rangle$$

since the qualities $r$ and $q$ are arbitrary, that $\langle r_k | r_i \rangle = \delta_{ki}$. Bra and ket are therefore "vectors" defined with respect to a "complete" orthonormalized basis.

Some remarks.

1) We have assumed for simplicity that $N(i)$, $M(i)$ are integer positive numbers. Actually, this condition may be removed. The quality $r$, for example, can be a physical quantity which assumes continuous values on a real segment (which can possibly be extended, with a passage to the limit, to the real straight line or to a half-straight line). This segment can thus be subdivided into $L$ identical contiguous segments and the index $i$ can be applied to distinguish these segments. In other words, one can let $r = r_i$ if the value of $r$ falls within the $i$-th segment. One can then thicken the subdivision of the domain of $r$ by making $L$ tend to infinity ($L \to \infty$). In this passage to the limit the numbers $N(i)$, $M(i)$ remain defined if they are re-interpreted as the <u>fraction</u> of transactions for which $r$ falls within the $i$-th interval. With this re-definition (implicit in the introduction of the normalization factor Z), they become real for every finite $L$ and remain real (infinitesimals) in the passage to the limit $L \to \infty$. The actual amplitude is the limit value of the amplitude for $L \to \infty$, provided that this limit exists.

Regarding the sign of $N(i)$, $M(i)$ it must be borne in mind that at the end of the procedure these numbers appear in the definition of the transaction amplitude. The probability of the transaction is given by the modulus squared of the amplitude and is therefore not affected by the multiplication of the amplitude by a phase factor $exp(i\delta)$. In particular, if $\delta = \pi$ this factor is equal to -1. Thus, by multiplying the amplitude by -1, the probability of the transaction remains unchanged. But to multiply the amplitude by -1 means to multiply the numbers $N(i)$, $M(i)$ which appear in its definition by -1, i.e. to redefine them as negative. Thus the condition that these quantities be positive or null can also be removed.

With this removal, the meaning of the numbers $N$, $M$ changes slightly. A positive value of $N$ signifies a *creation* of connections of the first group (Fig. 3a), while a negative value of $N$ signifies the *destruction* of an equivalent number of such connections. Similarly, a positive value of *M* signifies a *creation* of connections of the second group (Fig. 3b), while a negative value of *M* signifies the *destruction* of an equivalent number of such connections. The sum and product of these numbers becomes the sum and product of operations that create and destroy a certain type of connection.

2) The concept of probability implies that of randomness. There is nothing mysterious about quantum randomness. The transaction ( $r \mid q$ ) <u>is constituted</u> by its extremes which are the events $\mid q )( q \mid$ and $\mid r )( r \mid$, and is therefore completely defined once these two extremes are given. There is therefore no effective indetermination - once the initial and final conditions have been assigned the transaction is completely determined, because it actually only consists of the pair of these two conditions.

The problem arises when only one of the two conditions is given and one wants to know which the other one can be. The other condition is then determined by the physical reality bootstrap process which, as we have seen, comprises acausal elements; to predict this condition with certainty, therefore, is generally not possible.

One must therefore reason in statistical terms, introducing a probability of the transaction taking place. In the process in question, the vacuum is coupled with itself through a certain number of closed rings or loops. Each ring is broken into two half-rings whose extremes, *A* and *B*, are equal for all the rings in question; the events $A = \mid q )( q \mid$ and $B = \mid r )( r \mid$ are associated with these extremes. Two oppositely-oriented half-rings can reconnect to these extremes, giving rise to a new closed ring which comes to constitute the transaction. This reconnection can be done in $n^2$ ways, where *n* is the number of half-rings having a given orientation and belonging to a given group (that is, $n = N$ or $n = M$). All these ways are identical with respect to the bootstrap process and are therefore *a priori* equiprobable. The probabilistic reasoning described above now follows.

3) The transaction ( $s \mid q$ ) has been associated with the amplitude $\langle s \mid q \rangle$ with:

$$\mid q \rangle = \sum_k \mid r_k \rangle \langle r_k \mid q \rangle \quad .$$

The quality *q* also includes, in its definition, the instant $t(q)$ at which the event $\mid q )( q \mid$ occurs, so that the ket $\mid q \rangle$ depends on $t(q)$. The preceding equation thus implies that also the amplitudes $\langle r_k \mid q \rangle$ and kets $\mid r_k \rangle$ are dependent upon $t(q)$. Yet $\langle r_k \mid q \rangle$ is the amplitude of the ( $r_k \mid q$ ) transaction, in which two distinct time instants appear: that of the event $\mid q )( q \mid$ and that of the event $\mid r_k )( r_k \mid$; that is, $t(q)$ and $t(r)$, respectively. The dependence of the right-hand member of the equation on $t(q)$, $t(r)$ is compatible with the dependence of the left-hand member on $t(q)$ only if it is assumed that $t(q) = t(r)$. Therefore the kets $\mid q \rangle, \mid r_k \rangle$ and the bras $\langle r_k \mid$ must be considered at the same time instant.

The same reasoning applies to the expression:

$$\langle s \mid = \sum_i \langle s \mid r_i \rangle \langle r_i \mid \quad ,$$

for which it therefore turns out that $t(s) = t(r)$. On the other hand, the amplitude $\langle s \mid q \rangle$ can be constructed by left-multiplying the ket $\mid q \rangle$ by the bra $\langle s \mid$. This amplitude therefore makes sense only if the conditions $t(q) = t(r)$, $t(s) = t(r)$ are simultaneously satisfied; by the transitive property of equality, one thus obtains $t(s) = t(q)$. If $t(s)$ and $t(q)$ are different, then the amplitude $\langle s \mid q \rangle$ makes sense only if a transport rule exists for the ket $\mid q \rangle$ from $t = t(q)$ to $t = t(s)$, or a transport rule for the bra $\langle s |$ from $t = t(s)$ to $t = t(q)$. In the first case the ket $| q \rangle$ is transported to $t = t(s)$ and then is left-multiplied by $\langle s |$; in the second case the bra $\langle s |$ is transported to $t = t(q)$ and is then right-multiplied by $| q \rangle$. The transport rule is defined by the time evolution operator S, but the quantum dynamics will not be discussed here for reasons of space; interested readers are referred to other works (3). Strictly speaking, therefore, the portion of formalism developed here is only valid for the qualities q, r, s associated with kets or bras which do not vary over time (constants of motion), or which vary to a negligible degree within the interval $(t(q), t(s))$.

## 4.1 Quantum Formalism

Essentially, there is nothing mysterious about quantum formalism, provided that it is gradually and properly introduced. We have seen that a generic vector (ket or bra) can be decomposed into a complete basis of orthonormalized vectors. For example:

$$| \psi \rangle = \sum_i c_i | \varphi_i \rangle$$

$$\langle \varphi_k | \varphi_i \rangle = \delta_{ki},$$

where $c_i$ are complex numbers. The probability of the $(\varphi_k | \psi)$ transaction, therefore, is expressed by Born's rule, bearing in mind the caveats on the time variation of the quantities already discussed in the previous section:

$$P(\varphi_k | \psi) = |\langle \varphi_k | \psi \rangle|^2 / \langle \psi | \psi \rangle = c_k^* c_k / \sum_i c_i^* c_i.$$

Let us now suppose that a certain physical quantity O, in the realization of the $(\varphi_k | \psi)$ transaction, assumes the value $o(k)$. Let us assume that this value is a real number; this assumption does not constitute a loss of generality because it is practically always possible to satisfy it by adopting a suitable system of definitions.

Let us suppose that the quality $\psi$ is fixed, while the quality $\varphi$ can freely vary on its support set $\{\varphi_k\}$. Therefore the *a priori* probability that the outcome is quality $\varphi_k$ is $P(\varphi_k | \psi)$; the expectation value of O is given by:

$$\langle O \rangle_\psi = \sum_k P(\varphi_k | \psi) o(k) =$$

$$\sum_k o(k) c_k^* c_k / \sum_i c_i^* c_i.$$

This formula is entirely general. A very important special case is that in which an appropriate self-adjoint linear operator $\Omega$ exists on the rigged space of kets and bras, such that:

$$\Omega | \varphi_i \rangle = o(i) | \varphi_i \rangle \quad \text{for every value of } i.$$

From the self-adjointness of $\Omega$ follows that:

$$\langle \varphi_i | \Omega = \langle \varphi_i | o(i) \quad \text{for every value of } i.$$

It immediately follows from the linearity of Ω that:

$$\Omega |\psi\rangle = \sum_i c_i \, \Omega |\varphi_i\rangle = \sum_i c_i \, o(i) |\varphi_i\rangle \;;$$

$$\langle \psi |\Omega| \psi \rangle =$$

$$\left( \sum_k c_k^* \langle \varphi_k | \right) \left( \sum_i c_i \, o(i) |\varphi_i\rangle \right) = \sum_k o(k) \, c_k^* \, c_k \,.$$

And since:

$$\langle \psi | \psi \rangle = \left( \sum_k c_k^* \langle \varphi_k | \right) \left( \sum_i c_i |\varphi_i\rangle \right) = \sum_i c_i^* \, c_i \,,$$

one obtains:

$$\langle O \rangle_\psi = \langle \psi |\Omega| \psi \rangle / \langle \psi | \psi \rangle \,.$$

This formula is less general but frequently used in quantum physics. The operator Ω is thus said to be "associated" with the quantity $O$. The $o(k)$ values in this case form the "spectrum" of $O$ (or of Ω). The Hermitianity of Ω ensures on the one hand that the $o(k)$ values are real and on the other that the procedure is symmetrical. Indeed, we would have obtained the same result by developing $\langle \psi |$ Ω and then right-multiplying by $|\psi\rangle$. One ought to bear in mind that, given a physical quantity $O$, a self-adjoint linear operator associated with it does not always exist; a well-known example is the time $t$. The "postulate" set out in many text books, whereby to every $O$ there corresponds an Ω, is not correct. In many important cases, though, Ω exists and the aforementioned formula allows to draw advantage from this.

**4.2 Structure of Transactions**

It seems appropriate to elucidate a question which could easily be a source of misunderstanding. Let us consider a generic transaction $(\varphi | \psi)$ during which the quality $\psi$ is created and the quality $\varphi$ is destroyed. It is constituted by the two events $A = |\psi)(\psi|$ and $B = |\varphi)(\varphi|$, which take place at the instants $t_A$ and $t_B$, respectively. <u>These two events are the only manifest reality in the physical world.</u> The "facts" are constituted solely by these two events.

These two events, on the other hand, are correlated. The correlation between $A$ and $B$ is represented, as we have seen, by the "amplitude" of the transaction, $\langle \varphi | S | \psi \rangle$. This amplitude can generally be developed as the sum of the partial amplitudes corresponding to different paths joining $A$ and $B$. Each of these partial amplitudes can then be represented as the product of amplitudes of free propagations joining interaction vertices.

For example, let us consider the well-known double-slit experiment. In such a case, $|\psi\rangle$ is the ket associated with the "preparation" of the "particle" in the source and $\langle \varphi |$ is the bra associated with the "detection" of the "particle" behind the screen containing the slits.

Under the action of $S$, $|\psi\rangle$ evolves freely up to the interaction with the screen. If the screen is completely absorbent, then, indicating as $x$ the position of a generic point on its rear wall, one has $\langle x | \psi \rangle = 0$, unless $x$ corresponds to one of the two slits. The following time evolution of $| \psi \rangle$ is still free. Therefore, indicating the free time evolution operator of the particle as $S$ and the two slits as $f_1, f_2$, one can write:

$$\langle \varphi | S | \psi \rangle = \langle \varphi | S | f_1 \rangle \langle f_1 | S | \psi \rangle +$$

$$\langle \varphi | S | f_2 \rangle \langle f_2 | S | \psi \rangle \ .$$

As can be seen, one has the sum of two partial amplitudes, one of which corresponds to a "passage through slit $f_1$", the other to a "passage through slit $f_2$". These two amplitudes interfere, for the reasons seen above. Each of the two amplitudes is represented, in turn, as the product of two free propagation amplitudes of the "particle": one corresponding to the propagation from the source to the slit, the other from the slit to the detector.

Hence: either the "particle" interacts with the atoms which compose the screen, and is absorbed, or it does not interact with the screen at all. Only in this second case can it reach the detector, passing through $f_1$ or through $f_2$. The terms $f_1, f_2$ which appear in the expression of the amplitude are therefore associated with interactions with the atoms of the screen [to be precise, "negative" interactions, i.e. non-interactions or absence of actual interactions]. The correlation between events $A$ and $B$ is therefore conditioned by the atoms of the screen.

And what are the atoms of the screen? Nothing other, in fact, than events of the $| \Lambda ) ( \Lambda |$ type, where $\Lambda$ is the "atomic quality". Therefore, the correlation between events $A$ and $B$ is affected by the other events of the Universe, in this specific case by the events $| \Lambda ) ( \Lambda |$ which constitute the screen. The structure of $\langle \varphi | S | \psi \rangle$ includes these influences.

Now, the fundamental thing to understand is that *the interaction events which appear in $\langle \varphi | S | \psi \rangle$ are not real, in fact they are not events.* For example, the (non) interaction events $f_1, f_2$ in the example we have just described are not real events. By this we mean to say that these events are not pairs of emission/re-absorption of qualities by the physical vacuum, whereas the extreme events $A$, $B$ are.

One must be careful to consider, however, that the events $| \Lambda ) ( \Lambda |$ are real, too, and that the transaction amplitude actually expresses the influence of the other <u>real</u> events of the Universe, such as these very events $| \Lambda ) ( \Lambda |$, on the correlation between the <u>real</u> events $A$ and $B$.

This fact that the virtual events appearing in the partial amplitudes as mere artifices of calculation do not have a physical reality of their own should not lead to believe, however, that they bear no relation with physical reality. Indeed, as has been seen, they represent actual aspects of the transaction bootstrap from the vacuum: the partial amplitudes are associated with the half-loops. Furthermore, virtual events can become real. For example, a "particle" counter can be placed behind $f_1$ or $f_2$. In this case, though, one is no longer dealing with the original transaction, rather with other entirely different transactions, for example $(f_1 | \psi )$ [6].

The free propagations $\langle f_1 | S | \psi \rangle, \langle \varphi | S | f_1 \rangle, \langle \varphi | S | f_2 \rangle, \langle f_2 | S | \psi \rangle$ can be represented, as is well known, in the form of the propagator of a particle of energy $E$, impulse $\boldsymbol{p}$ and spin $\boldsymbol{s}$ (and given internal quantum numbers): $\langle E, \boldsymbol{p}, \boldsymbol{s} | S | E, \boldsymbol{p}, \boldsymbol{s} \rangle$. The "particle" therefore is actually nothing other than a packet of constants of motion which are propagated, and does not exist as substantial reality.

In this respect, an issue which has given rise to quite a number of misunderstandings is that which concerns the sign of energy $E$. Energy $E$ appears from the vacuum at the time of creation, and vanishes into the vacuum at the time of destruction, as do, in fact, the other quantities which make up the quantum.

---

[6] The event $| f_1 ) ( f_1 |$ is real, whether the "particle" is detected behind $f_1$, or is not detected behind $f_2$, if third possibilities are excluded. In this second case, one has a null interaction vertex.

From a merely numeric standpoint, one can view creation as the release by the vacuum of an $E$ amount of energy to the manifest physical world. Destruction will thus be the release by the manifest physical world of an amount of energy $E$ to the vacuum. Now, it is known that a system's exchanges of energy with the "outside" must be taken with a different sign according to their direction. We may therefore adopt two conventions.

First convention. We may assume that the energy released by the vacuum is positive and that absorbed by the vacuum is negative. In this case, we shall have at creation the exchange of an energy $E > 0$, and at destruction the exchange of an energy $-E < 0$.

Second convention. We may assume that the energy released by the vacuum is negative and that absorbed by the vacuum is positive. In this case, we shall have the exchange of an energy $E < 0$ at creation, and the exchange of an energy $-E > 0$ at destruction.

Now, any transaction falls within a chain of transactions unlimited in the two directions of time and to which a total order relation applies, which is chronological ordering. The direction of such an ordering must be the same for all the interacting chains which make up the physical world, if one is to have a global time coordinate that is definable. Once the direction of the time ordering has been chosen, creation precedes destruction, based on the convention that creation is called the extreme of the transaction which occurs first.

Thus, with the first convention we have the creation of an energy $E > 0$ at $t_0$ and the absorption of an energy $E < 0$ at $t_1 > t_0$. This corresponds to the forward propagation in time of a quantum of energy $E > 0$, or to the propagation backwards in time of a quantum of energy $E < 0$. Indeed the creation of a quantum of energy $E < 0$ at $t_1$ is equivalent to the destruction of a quantum $E > 0$ and the destruction of a quantum $E < 0$ at $t_0$ is equivalent to the creation of a quantum of energy $E > 0$.

With the second convention we have the creation of an energy $E < 0$ at $t_0$ and the absorption of an energy $E > 0$ at $t_1 > t_0$. This corresponds to the forward propagation in time of a quantum of energy $E < 0$, or to the propagation backwards in time of a quantum of energy $E > 0$. Indeed, the creation of a quantum of energy $E > 0$ at $t_1$ is equivalent to the destruction of a quantum $E < 0$ and the destruction of a quantum $E > 0$ at $t_0$ is equivalent to the creation of a quantum of energy $E < 0$.

Clearly the two conventions, though equally possible, are incompatible: either one or the other must be chosen. If the first is chosen, as is usual, retropropagations in time of quanta with positive energy cannot exist[7]. One thus has a "time arrow" of entirely microphysical origin and which is absolutely elementary. This concept can also be applied to the solutions of any wave equation, for example the D'Alembert wave equation with a four-current term in the right-hand. This entails the non-existence of advanced potentials; indeed, they do constitute valid solutions, but under the opposite convention.

The existence of a time arrow also in a mechanism which is of itself totally symmetrical in time, such as a transaction, is a fact of capital importance. It implies the causality of quantum phenomena, and daily proof that a cause always precedes its effect. Time, therefore, flows in one direction only, even if the transaction - the elemental process of the physical world - is symmetrical in time.

In discussing the double slit experiment we considered a situation in which two events $A$ and $B$ are connected by an operator $S$ which can be represented as the sum of partial propagation amplitudes of *individual* quanta. This situation is typical of microphysics experiments, but not of everyday reality. In everyday life one deals with transaction aggregates, rather than with individual transactions. A piece of "common matter" is actually an aggregate of transactions which take place in enormous numbers per second and per cubic centimeter. It is absolutely impossible, in a typical situation, to follow the alternation of individual transactions, i.e. to do that which in technical jargon is known as "following the time evolution of the system microstate". One therefore confines oneself to taking into consideration certain *global* qualities which in the course of time evolution are preserved or show a sufficiently "continuous" trend.

---

[7] Should this occur, one is actually considering a positive energy propagation, forward in time, of the conjugated of the quantum in question (CPT theorem).

Thus the idea is born of a persistent "object", which constitutes the basis of classical physics and which instead has no part to play at the individual transaction level.

## 4.3 Wavefunctions

The general quantum formalism introduced in the preceding sections, and which basically constitutes the Dirac formalism (30) admits, as a special case, of a less general formalism known as the "wavefunction" formalism.

Let us consider the transaction $(q \mid \psi)$ during which the quality $\psi$ is created at the time $t_1$ and the quality $q$ is destroyed at the time $t_2 > t_1$. We assume that the quality $q$ can be represented by a set of $N$ values of an equal number of real variables $q_i$ ($i = 1, 2, ..., N$), called "positions" or "generalized coordinates". The amplitude associated with this transaction is $\langle q \mid S(t_2, t_1) \mid \psi \rangle$.

In the limit $t_2 \to t_1$ one has $S(t_2, t_1) \to 1$, and the amplitude tends to $\langle q \mid \psi \rangle = \psi(q)$. In general this limit amplitude depends on $t = t_1 = t_2$, and it is indicated, therefore, with $\psi(q, t)$. This is the "wavefunction". Wavefunctions are therefore a special class of amplitudes and their formalism can be obtained from the Dirac general formalism by particularization. It is possible, however, to introduce the quantum formalism by reasoning directly on the wavefunctions. This approach is less general than the one seen above, but is of considerable theoretical interest, so we propose to describe it here.

The wavefunction concept is useful when there exists a real function $A(q, t)$, called "action", of the "Lagrange coordinates" $q$ and of the time $t$ such that the "generalized pulses" $p_i = \partial A(q, t)/\partial q_i$ are constants of motion in the free evolution.

The function $\psi(q, t)$ is assumed to be decomposable, by means of a Fourier transform, into the sum of harmonic components (letting $\hbar = h/2\pi = 1$):

$$\phi(p) \, exp(i \, p \bullet q \, - i \, Et) \quad .$$

On thus obtains:

$$-i(\partial/\partial q_i) \, [\phi(p) \, exp(i \, p \bullet q \, - i \, Et)] \; = \; p_i \, [\phi(p) \, exp(i \, p \bullet q \, - i \, Et)] \quad .$$

It follows from this relation that: a) the operator associated with $p_i$ is $-i(\partial/\partial q_i)$ ; b) the harmonic components are eigenfunctions of this operator. Therefore $p$ assumes a definite value on these components (that is, its fluctuations are null). In addition, one obtains:

$$i(\partial/\partial t) \, [\phi(p) \, exp(i \, p \bullet q \, - i \, Et)] \; = \; E \, [\phi(p) \, exp(i \, p \bullet q \, - i \, Et)] \quad .$$

That is: a) the operator associated with $E$ is $i(\partial/\partial t)$; b) the harmonic components are eigenfunctions of this operator. Therefore $E$ assumes a definite value on these components (that is, the fluctuations of $E$ are null). That $E$ is the energy, i.e. an eigenvalue of the Hamiltonian $H$, is easily seen by using the very definition of Hamiltonian as an exponent in the evolution operator. If one takes $\phi(p) \, exp(i \, p \bullet q)$ as the initial wavefunction, one must have:

$$\phi(p) \, exp(i \, p \bullet q \, - i \, Et) \; = \; exp(-i \, Ht) \, \phi(p) \, exp(i \, p \bullet q) \quad .$$

This relation implies $H = E \cdot$, i.e. the fact that the harmonic component is an eigenfunction of $H$ with an eigenvalue $E$. On the other hand, the operator associated with $E$ is $i(\partial/\partial t)$, so that in conclusion $H = i(\partial/\partial t)$.

Classically, the Hamiltonian function depends on $q$ and on $p$. A common, in many cases useful, empirical method for constructing the Hamiltonian operator is to substitute $p_i$ in the classical expression with $-i(\partial/\partial q_i)$. One can thus write in symbols $H = H(q, -i\partial_q)$. Assuming that:

$$\psi(q, t) = \sum_p \phi(p) \exp(i\, p \bullet q - i\, Et)$$

One has, given the linearity of the operators $H$ and $i(\partial/\partial t)$:

$$[H - i(\partial/\partial t)]\, \psi(q, t) = \sum_p \phi(p)\, [H - i(\partial/\partial t)]\, \exp(i\, p \bullet q - i\, Et) =$$

$$= \sum_p \phi(p)\, [H - E]\, \exp(i\, p \bullet q - i\, Et) = 0 .$$

I.e. the equation of motion:

$$H(q, -i\partial_q)\, \psi(q, t) = i(\partial/\partial t)\, \psi(q, t) .$$

This equation defines the time evolution of the $\psi$. It is fundamental to understand that the $q$, on which the $\psi$ depends, are not spatial coordinates but generalized coordinates introduced on the assumption that an action exists for the process being studied. As is well known, the Lagrange coordinates coincide with the spatial coordinates in one case only: that of the classical "material point". Therefore the function $\psi$ does not represent an actual physical field on spacetime.

To understand the actual physical meaning of the wavefunction, one must consider its role in a defined transaction process.

Let us firstly consider a transaction $(E', p' \mid E, p)$, in which at the time $t_1$ a harmonic component is created which has a moment $p$ and energy $E$, and at the time $t_2 \geq t_1$ a harmonic component is destroyed which has a moment $p'$ and energy $E'$. In the limit $t_2 \to t_1$, the amplitude of this transaction is

$$\langle p' \mid p \rangle \langle E' \mid E \rangle = \delta(p - p')\, \delta(E - E') .$$

Therefore, in the instantaneous limit, the amplitude of the transaction is equal to 1 if $p = p'$ and $E = E'$, otherwise it is 0. In other words, in this limit the interaction effect disappears, so that $p$ and $E$ are preserved. The meaning of this formalism is therefore the following (spin is disregarded and it is assumed that $E > 0$):

    a) A quantum $(E, p)$ is emitted at $t_1 = t$.
    b) A quantum $(E', p')$ is absorbed at $t_2 = t + \varepsilon$. $\varepsilon \to 0$, $\varepsilon > 0$.
    c) The emission and absorption of the quantum are possible only if $p = p'$ and $E = E'$.
    d) In this case, the number of quanta emitted and absorbed (i.e. 1) is given by the amplitude (which is indeed equal to 1).

    e) In the same transaction, a quantum $(-E', -p')$ is emitted at $t_2 = t + \varepsilon$.

f) In the same transaction, a quantum ($-E$, $-p$) is absorbed at $t_1 = t$.

g) The emission and absorption of the quantum are possible only if $p = p'$, $E = E'$.

h) In this case, the number of quanta emitted and absorbed (i.e. 1) is given by the complex conjugate amplitude of the preceding one (which is indeed equal to $1^* = 1$).

The quanta propagating forward in time are the half-loops, pointing towards the future, which connect the two transaction events. The quanta propagating back in time are the half-loops, pointing towards the past, which connect the two transaction events. The number of complete loops obtained by the rejoining of the two half-loops, one pointing to the future and the other pointing towards the past, is therefore equal to the number of quanta propagated forward in time by the number of quanta propagated backwards in time. In this specific case, the number of closed loops is 1 x 1 = 1. Generally speaking, this number is proportional to the transaction probability.

Let us now turn to the general case of a transaction ($\phi \mid \psi$) in the course of which the quality $\psi$ is created at the time $t_1$, and the quality $\phi$ is destroyed at the time $t_2 \geq t_1$. The amplitude of the process forward in time is $\langle \phi \mid S(t_2, t_1) \mid \psi \rangle$, and the amplitude of the process backwards in time is the complex conjugate of the latter. Letting:

$$\langle q \mid S(t_2, t_1) \mid \psi \rangle = \frac{1}{(2\pi)^{N/2}} \int dp\, e^{i\, p \bullet q}\, \alpha(p)$$

$$\langle \phi \mid q \rangle = \frac{1}{(2\pi)^{N/2}} \int dp'\, e^{-i\, p' \bullet q}\, \beta^*(p') \quad ,$$

the forward amplitude is expressed by:

$$\frac{1}{(2\pi)^N} \int dp\, dp'\, \alpha(p) \beta^*(p') \int dq\, e^{i\, p \bullet q}\, e^{-i\, p' \bullet q}$$

$$= \int dp\, dp'\, \alpha(p) \beta^*(p') \delta(p - p') \quad = \quad \int dp\, \alpha(p) \beta^*(p)$$

which is equal to a certain complex number $\Gamma = \Gamma_0\, exp(i\delta)$, where $\Gamma_0$ is real and positive. The amplitude backwards in time is the complex conjugate $\Gamma^* = \Gamma_0\, exp(-i\delta)$ and the number of closed loops reconstituted by the joining of the half-loops is $\Gamma^*\, \Gamma$, real and positive. This number is proportional to the transaction probability.

One could object that $\Gamma$, being a generic complex number, cannot be the number of half-loops - which obviously must be either integer and positive or real and positive. On must note, however, that by multiplying the two wavefunctions $\alpha$ and $\beta$ under the integral sign by $exp(i\eta)$, $exp(-i\varphi)$, respectively, with $\eta - \varphi = -\delta$, the phase factor $exp(i\delta)$ in $\Gamma$ is cancelled out and only $\Gamma_0$ survives, real and positive. The number of loops is therefore given by $\Gamma_0^*\, \Gamma_0$, which is still equal to $\Gamma^*\, \Gamma$. It follows from the Schrödinger equation, on the other hand, that multiplying the wavefunction by a constant phase factor does not entail any change in the dynamics. Therefore, multiplying the wavefunctions by the phase factors $exp(i\eta)$, $exp(-i\varphi)$ does not have any physical effects.

The fact that the transaction probability turns out to be proportional to $|\Gamma|^2$ constitutes Born's rule. In the expression shown, the integral in $q$ corresponds to the count of a single emitted-absorbed quantum. The double integral in $p$ and $p'$ corresponds to the sum of these individual counts, i.e. to the half-loop count.

The construction of the quantum formalism proceeds in an entirely parallel fashion to that seen with bras and kets. If it is possible to associate with a physical quantity $O$, having a spectrum $\{ o(i) \, ; \, i = 1, 2, ... \}$, a self-adjoint linear operator $\Omega$ such that:

$$\Omega \phi_i(q) = o(i) \phi_i(q) \qquad i = 1, 2, ...$$

then it is also possible, by the well-known spectral theorem on self-adjoint linear operators, to orthonormalize the basis:

$$\int dq \, \phi_i(q) \phi_k^*(q) = \delta_{ik} \; .$$

Assuming therefore that:

$$\psi(q) = \sum_n \alpha_n \phi_n(q) \quad ,$$

the amplitude of the $(\phi_k | \psi)$ transaction in the instantaneous limit is expressed by:

$$\frac{\int dq \, \psi(q) \phi_k^*(q)}{\left[ \int dq \, \psi(q) \psi^*(q) \right]^{1/2}} = \alpha_k / \left( \sum_n |\alpha_n|^2 \right)^{1/2} \; .$$

And by normalizing the $\psi$ to 1, one thus has, for the transaction probability, the value $|\alpha_k|^2$. Therefore if $\psi$ is the fixed extreme of the transaction and $\phi_k$ is the free extreme, the value $o(k)$ of the quantity $O$ will be realized with a probability of $|\alpha_k|^2$. The average value of the quantity $O$ on the ensemble of all the $(\phi_n | \psi)$ transactions, where $n$ is variable, is therefore:

$$\sum_n o(n) |\alpha_n|^2 = \int dq \, \psi^*(q) \Omega \psi(q) \quad ,$$

as can be verified immediately- This is the rule for calculating $<\Omega>_\psi$.

## 5. TRANSACTIONAL MEANING OF THE WICK ROTATION

The formalism of wavefunctions is especially appropriate for approaching, from a transactional point of view, one of the most time-honoured mysteries of theoretical physics: the connection between time and thermodynamics mediated by the Wick rotation (31).

**5.1 Quantum Leap in the Energy Representation**

Let us consider the transactional loop:

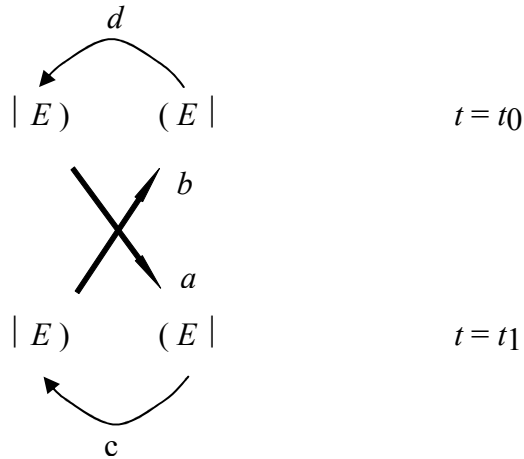

The vacuum is connected to itself through the *acbd* cycle or through the inverse cycle *dbca*. The two events $\mid E\,)(\,E\mid$ which occur at the instants $t_0$ and $t_1 > t_0$ represent the emission and the absorption, respectively, of a quantum (electron, photon, muon, etc.) of exactly defined energy equal to $E$.

The hypothesis is that the entire process - say, *acbd* - takes place as a virtual self-coupling process of the archaic vacuum. If, at the time instants $t_0$ and $t_1$, the coupling is possible with other transactional loops - that is, if the events $\mid E\,)(\,E\mid$ *actually exist* - the virtual process becomes real and one has the unfolding of a transaction over spacetime. This is how the network of events over spacetime which forms the observed material Universe emerges. Since the coupling satisfies the principles of conservation, macroscopic causality is complied with.

In the absence of the events $\mid E\,)(\,E\mid$, the virtual process *acbd* is an indivisible whole; the entire cycle must be covered in order to return to the point of departure, i.e. to the vacuum. From the point of view of the vacuum, therefore, this cycle is an elementary event. We can hypothesize that an occurrence probability is associated with this elementary event:

$$P = \exp(-2E/kT) ,$$

with $T > 0$. We can divide the cycle into the two processes *da*, *cb* which are not true events (they cannot occur by themselves) and factorize $P$ according to the symmetrical expression:

$$P = P_{da} \times P_{cb}$$

$$P_{da} = P_{cb} = \exp(-E/kT) .$$

Obviously, the factors $P_{da}$, $P_{cb}$ are not probabilities because, as we have said, the processes *da*, *cb* are not events.

When the spacetime unfolding of the transaction occurs, the half-processes *da*, *cb* become true distinct processes on spacetime, instead. Process *da* consists of event $\mid E\,)(\,E\mid$ which occurs at $t = t_0$ and of the forward connection with the event $\mid E\,)(\,E\mid$ which occurs at $t = t_1$. Process *cb* consists of the event $\mid E\,)(\,E\mid$ which occurs at $t = t_1$ and of the backward connection with event $\mid E\,)(\,E\mid$ which occurs at $t = t_0$. We are dealing therefore with the two sub-processes which constitute the transaction.

The spacetime unfolding corresponds to a Wick rotation. For the first process the Wick rotation takes the form:

$$1/kT \rightarrow -i(t_1 - t_0)/\hbar \ ,$$

while for the second process it takes the form:

$$1/kT \rightarrow -i(t_0 - t_1)/\hbar \ .$$

In these equations, the variable $t$ indicates the proper time of the quantum of energy $E$ exchanged between the two events. Consequently, the factors $P_{da}$, $P_{cb}$ become:

$$P_{da} \rightarrow \Pi_{da} = exp[iE(t_1 - t_0)/\hbar]$$

$$P_{cb} \rightarrow \Pi_{cb} = exp[-iE(t_1 - t_0)/\hbar] \ .$$

Thus, the product of these factors is the transform of $P$ according to the Wick rotation; this product is equal to 1. It is still a probability, but its meaning now is entirely different. This is the probability that, given the emission (absorption) of a quantum of energy $E$, the energy of the successively absorbed (previously emitted) quantum is $E$. Since the propagation of the quantum is free (we are considering a single transaction without intermediate vertices) this probability is certainly 1. Alternatively, one can say that 1 is the number of quanta of energy $E$ propagated between the emission and absorption events.

More generally, let us suppose that an action variable $S$ exists such that:

$$E = - \partial_t S \ .$$

If $E$ is a constant of motion, as in the example being considered, then:

$$E = - \Delta S/\Delta t$$

and therefore:

$$P_{da} = exp(-E/kT) = exp[-\Delta S/(\Delta t \, kT)] \rightarrow exp[-(\Delta S/\Delta t)(-i\Delta t/\hbar)]$$

$$= exp(i\Delta S/\hbar) \ ;$$

$$P_{cb} = exp(-E/kT) = exp[-\Delta S/(\Delta t \, kT)] \rightarrow exp[-(\Delta S/\Delta t)(i\Delta t/\hbar)]$$

$$= exp(-i\Delta S/\hbar) \ ;$$

where $\Delta t = t_1 - t_0$. One must note that the last equalities obtained remain valid even if the quantum mass is null. They are also relativistically invariant, and are therefore valid in a general frame of reference in which $\Delta S = E\Delta t - \mathbf{p}\Delta\mathbf{x}$, where the vector $\mathbf{x}$ represents the generalized coordinates of the quantum and the vector $\mathbf{p}$ represents its conjugate pulses.

In conclusion, the Wick rotation connects the probability of a virtual process to the probability of the same process once it has become real.

## 5.2 Quantum Statistics in the Energy Representation

Let us now consider the case in which the event at $t = t_1$ actually were $|E')(E'|$ with $E' \neq E$. In this case, the coupling would not have been possible, as this would have violated the principle of the conservation of energy. The probability of this transaction would thus have been zero. Since, in this case,

$$P_{cb} \rightarrow exp[-iE'(t_1 - t_0)/\hbar] \ ,$$

the two cases are combined into one, assuming that the probability of the transaction is expressed by:

$$(1/2\pi\hbar)\int exp[iE(t_1 - t_0)/\hbar] \, exp[-iE'(t_1 - t_0)/\hbar] \, d(t_1 - t_0) \ = \ \delta(E - E') \ .$$

This expression can be written:

$$\int \psi\varphi^* \, d(t_1 - t_0) \ = \ \delta(E - E') \ ,$$

letting:

$$\psi \ = \ (1/2\pi\hbar)^{-1/2} \, \Pi_{da} \ , \qquad \varphi \ = \ (1/2\pi\hbar)^{-1/2} \, (\Pi_{cb})^* \ .$$

Let us now consider a more general transaction connecting the event $|\psi)(\psi|$ which occurs at $t = t_0$ with the event $|\psi)(\psi|$ which occurs at $t = t_1$. Let us suppose that the $\psi$ is represented, in this case, by the linear superposition:

$$\psi \ = \ (1/2\pi\hbar)^{-1/2} \sum_{E} a_E \, exp[iE(t_1 - t_0)/\hbar] \ .$$

Then $\varphi = \psi$ and the probability of the transaction is expressed by the integral:

$$(1/2\pi\hbar)\int \sum_{E, E'} a_E a_{E'}^* \, exp[iE(t_1 - t_0)/\hbar] \, exp[-iE'(t_1 - t_0)/\hbar] \, d(t_1 - t_0) \ =$$

$$= \sum_{E, E'} a_E a_{E'}^* \, \delta(E - E') \ = \ \sum_{E} a_E a_E^* \ .$$

As can be seen, the number of quanta of energy $E$ exchanged in the course of this transaction is $a_E a_E^*$, and they represent a fraction:

$$a_E a_E^* / \sum_{E} a_E a_E^*$$

of the total number of quanta exchanged. In other words, the probability of emitting/absorbing a quantum of energy $E$ is expressed by the fraction:

$$a_E a_E^* / \sum_{E} a_E a_E^* \ = \ |\int \psi\varsigma^* \, d(t_1 - t_0) \, |^2 \ / \int \psi\psi^* \, d(t_1 - t_0)$$

where:

$$\xi = exp[iE(t_1 - t_0)/\hbar] .$$

In the transaction considered above, the initial "state" $\psi$ (emission) and the final "state" $\psi$ (absorption) coincide. Let us now suppose that instead of the event $|\psi)(\psi|$ which occurs at $t = t_1$ there is an event $|\varphi)(\varphi|$; let us suppose that the $\varphi$ is represented, in this case, by the linear superposition:

$$\varphi = (1/2\pi\hbar)^{-1/2} \sum_E b_E \, exp[iE(t_1 - t_0)/\hbar] .$$

For simplicity, let us also suppose that $\psi$ and $\varphi$ are normalized to 1. The previous formula for the transaction probability is thus generalized:

$$\left| \int \psi\varphi^* \, d(t_1 - t_0) \right|^2 .$$

This result constitutes Born's rule and its validity extends beyond the energy representation ($E$-representation), as has been shown in the previous sections.

### 5.3 Wick Rotation and Uncertainty Principle

Let us now return to the expression of the Wick rotation; to fix the concept in our minds, we shall consider the "forward" rotation:

$$1/kT \rightarrow -i(t_1 - t_0)/\hbar ,$$

but the line of reasoning is also applicable to the "backward" rotation obtained from this one by exchanging $t_1$ and $t_0$. By comparing the expressions:

$$P_{da} = exp(-E/kT) \quad \text{and} \quad \Pi_{da} = exp(i\Delta S/\hbar)$$

it can be seen the uncertainty principle $\Delta S \geq \hbar$ takes the "non-rotated" form $E \geq kT$. The meaning is obvious: energy exchanges of smaller amount than the thermal energy $kT$ cannot be distinguished from the "substratum" of thermal fluctuations within the thermostat and are therefore "virtual".

### 5.4 Action as Information

By inverting the relation:

$$P_{da} = P_{cb} = exp(-E/kT)$$

one obtains:

$$E = -kT \ln(\rho)$$

where $\rho = P_{da}$ , $P_{cb}$ . We can introduce a "number of cases" variable equal to $W = 1/\rho$ and an entropy:

$\Lambda = k \, lnW = k \, ln(1/\rho)$ ,

so that the obtained relation becomes $E = \Lambda T$. Therefore, according to the Wick rotation the quantity:

$E/kT = \Lambda/k$

is transformed into the quantity $\pm i\Delta S/\hbar$. In other words, the action is the "rotated" expression of entropy. This equivalence between action and entropy had already been evidenced by De Broglie (32). We also note that an action value of $\Delta S = \hbar \, ln(2)$ corresponds to a binary choice ($\rho = ½$). One can therefore measure the action adimensionally, as an amount of information (in bits):

$\Sigma = \Delta S/\hbar \, ln(2)$ .

It is clear from the above that quantum statistics in $E$-representation is *de facto* equivalent to the statistics of a canonical ensemble with a thermal bath at the absolute temperature $T$. It is this fact which led De Broglie to conjecture the presence of a "hidden thermostat" (32).
It must be noted, however, that the representation of the thermal bath in the form of an aether of particles having chaotic motion, located on spacetime and with which real particles supposedly interact (32), is absolutely naive. Such an approach views real particles as persistent objects which describe a continuous - though non-differentiable - trajectory in spacetime. Actually, the "thermal" representation outlined here considers the elementary quantum process of emission-absorption (transaction) as a whole and it is therefore nonlocal. Time, in particular, emerges from the archaic physical vacuum as a consequence of the unfolding described by means of the Wick rotation, so that the "thermostat" does not live in time but rather belongs to a timeless dimension which unfolds through the rotation. Any attempt to re-obtain locality through a completion of this representation with hidden variables is therefore doomed to failure.

## 6. CONCLUSIONS

In our opinion, the reformulation of the transaction concept presented in this work maintains the essential premises of Cramer's approach. Firstly, the extremes of a transaction are events of creation and destruction of certain physical qualities. In particular circumstances, these qualities can be mapped on the values assumed by a maximal complete set of compatible observables which are constants of motion, and in this case one is dealing with the propagation of a *quantum*. This is a consequential generalization of Cramer's idea, which indeed identifies such extremes with emission/absorption events in the context of a classical (time-symmetric) theory of fields. In addition, the reformulation presented is fully consistent with the current formalism, if **R** processes are identified with *actual* - and not purely virtual - interaction processes (*objective* reduction postulate).
What for Cramer is the complex consisting of offer and confirmation waves here becomes, more simply and consistently with the standard formalism, the set of the two "forward" and "backward" amplitudes of the elementary quantum process, associated with the evolution operators $S$ and $S^+$, respectively.
In this formulation, none of the paradoxes raised against TI appears. For example, Maudlin's argument (9) is unfounded because the change in the experimental setup brought about by moving detector $B$, though triggered by a signal coming from detector $A$, leads to a completely different transaction. In the first case, the microevent $B^+B^-$ is never produced, in the second case it is certainly produced. Since one of the two extreme microevents of the transaction changes, the transaction itself changes and therefore, in accordance with the arguments set out above, the time evolution operator of this process changes. We note, on the other hand, that the null signal from *A*, i.e. the event (not *A*)$^+$(not *A*)$^-$ is itself a transaction termination (null interaction). The second transaction assumes as its input state the output state of the first.

For the same basic reasons, no problem can arise with Afshar's experiment (33).

If the proposed reformulation appears to be an improvement for the purpose of understanding the elementary quantum process, no less interesting are the possibilities for its application to sectors of physics research still being debated. For example, the Wick rotation becomes, in a physically clear way, a rule for the unfolding of the transactional loop from the *implicate order*; this agrees with the approach suggested by Bohm. We are dealing here with a connection between synchronic and diachronic realms of the physical world, which is also an emergence of real phenomena from a background of virtual possibilities, and an emergence of *time*. An application to the problem of initial conditions in cosmology has recently been proposed (34).

From a more speculative perspective, it is possible that the network of transactions, defined consistently with the proposal summarized here, constitutes the causal aspect of a more general modality for connections between microevents, which we may call *archetypal*. For a definition of archetypes and their possible role in natural history, see ref. (35); in ref. (36) the scheme is described of a class of experiments addressing their possible detection.

It must be noted, finally, that the spirit of the proposal, especially as regards the reference to a background outside of space and time, can also be found in the so-called "possibilist TI" formulated by R.E. Kastner in her most recent works (37), (38).

ACKNOWLEDGMENTS

This work has been written at the kind invitation of Ruth Elinor Kastner, whom I thank wholeheartedly for her encouragement and the confidence she has placed in me.